\documentclass[prb,twocolumn,showpacs,superscriptaddress,floatfix]{revtex4}
\usepackage{graphicx,amsfonts,amssymb,amsmath,wasysym}

\usepackage{hyperref}

\usepackage{color}

\definecolor{g}{rgb}{.1,0.4,.1}
\definecolor{b}{rgb}{0,0.2,1}
\definecolor{rouge}{rgb}{0.82,0.,0.}
\definecolor{vert}{rgb}{0.,0.82,0.}
\definecolor{bleu}{rgb}{0.,0.,0.69}
\definecolor{m}{rgb}{0.82,0.,0.82}
\definecolor{vert2}{rgb}{0.,0.5,0.}
\definecolor{rougeclair}{rgb}{1.0,0.7,0.7}

\newif\ifhyper
\hypertrue
\ifhyper
\hypersetup{
 citecolor = {green},
 colorlinks = {true},
 urlcolor = {blue} 
}
\fi

\newcommand{\anb}{b^{\phantom\dagger}}
\newcommand{\crb}{b^\dagger}
\newcommand{\bn}{{\boldsymbol{n}}}
\newcommand{\bi}{{\boldsymbol{i}}}
\newcommand{\bj}{{\boldsymbol{j}}}
\newcommand{\bk}{{\boldsymbol{k}}}

\newcommand{\ws}{{\widetilde{\sigma}}}

\begin{document}

\title{Perturbative study of the Kitaev model with spontaneous time-reversal
  symmetry breaking}

\author{S\'ebastien Dusuel}
\email{sdusuel@gmail.com}
\affiliation{Lyc\'ee Louis Thuillier, 70 Boulevard de Saint Quentin,
80098 Amiens Cedex 3, France}

\author{Kai Phillip Schmidt}
\email{schmidt@fkt.physik.uni-dortmund.de}
\affiliation{Lehrstuhl f\"ur theoretische Physik, Otto-Hahn-Stra\ss e 4, D-44221
Dortmund, Germany}

\author{Julien Vidal}
\email{vidal@lptmc.jussieu.fr}
\affiliation{Laboratoire de Physique Th\'eorique de la Mati\`ere Condens\'ee,
CNRS UMR 7600, Universit\'e Pierre et Marie Curie, 4 Place Jussieu, 75252
Paris Cedex 05, France}

\author{Rosa Letizia Zaffino}
\email{zaffino@lptmc.jussieu.fr}
\affiliation{Laboratoire de Physique Th\'eorique de la Mati\`ere Condens\'ee,
CNRS UMR 7600, Universit\'e Pierre et Marie Curie, 4 Place Jussieu, 75252
Paris Cedex 05, France}


\begin{abstract}

We analyze the Kitaev model on the triangle-honeycomb lattice whose ground state
has recently been shown to be a chiral spin liquid. We consider two perturbative
expansions~: the isolated-dimer limit containing Abelian anyons and the
isolated-triangle limit. In the former case, we derive the low-energy effective
theory and discuss the role played by multi-plaquette interactions. In this
phase, we also compute the spin-spin correlation functions for any vortex
configuration.
In the isolated-triangle limit, we show that the effective theory is, at lowest
nontrivial order, the Kitaev honeycomb model at the isotropic point. We also
compute the next-order correction which opens a gap and yields non-Abelian
anyons.
 
\end{abstract}


\maketitle

%
%
\section{Introduction}
%
%

In two dimensions, particles may obey nontrivial braiding statistics
\cite{Leinaas77,Wilczek82_1}. However, a direct observation of these so-called
anyons remains one of the most challenging topics in physics.
Several good candidates have emerged in the last years among which the
fractional quantum Hall effect \cite{Camino05_2} but the braiding of Laughlin
quasi-particles has still not been performed despite recent proposals based on
Mach-Zehnder interferometer \cite{Feldman07}.

Interestingly, at a theoretical level, such exotic excitations may also arise in
spin systems \cite{Wen03,Kitaev03}. A simple example is provided by the
so-called toric code whose elementary excitations are known to behave as semions
\cite{Kitaev03}. Nevertheless, this model is difficult to implement because it
is based on four-spin interactions which are not easily reproduced in
experimental set-ups.
A better candidate is undoubtedly the Kitaev honeycomb model \cite{Kitaev06}
which involves only two-spin interactions. Indeed, such a system may be realized
experimentally in optical lattices either with cold atoms
\cite{Duan03,Jiang08,Aguado08} or with polar molecules
\cite{Micheli06}. Furthermore, in a suitable parameter range, a perturbative
low-energy effective model of the honeycomb model is the toric code
\cite{Kitaev03} extended with multi-anyon interactions \cite{Schmidt08}. One
must however keep in mind that, in the honeycomb model, one also has fermionic
excitations which have to be taken into account when braiding anyons
\cite{Dusuel08_1}.
The honeycomb model has attracted much attention recently
\cite{Lahtinen07,Schmidt08,Chen08,Sengupta08} because it can additionally be
solved exactly via different fermionization methods (Majorana
fermions\cite{Kitaev06} or Jordan-Wigner transformations\cite{Chen07_1}).

One of the most interesting extensions of this model suggested in Kitaev's
seminal paper \cite{Kitaev06} has been proposed by Yao and Kivelson \cite{Yao07}
who considered the same kind of model but on the triangle-honeycomb
lattice. Indeed, in the presence of odd cycles, the system spontaneously breaks
the time-reversal symmetry and has two topologically distinct gapped phases
characterized by Abelian and non-Abelian excitations.
In their study, Yao and Kivelson showed that the ground state is a
chiral spin liquid associated to an odd Chern number (note that such an exotic
state of matter has also been found in another spin model \cite{Schroeter07}).
Their whole analysis relies on an exact treatment of the vortex-free sector (see
below for details) which allows them to compute the fermionic gap. 
However, as in the honeycomb model, although the ground state belongs to this
subspace, the low-energy states are known to lie in other vortex sectors for a
wide range of parameters in the Abelian phase. In contrast, near the transition
point, the fermionic gap in the vortex-free sector is smaller than the vortex
gap \cite{Yao07}. The aim of this paper is to analyze this low-energy spectrum
following and extending the procedure developed in Ref.~\onlinecite{Schmidt08}
for the honeycomb model.

This paper is organized as follows. In Sec.~\ref{sec:model}, we introduce the
Kitaev model on the triangle-honeycomb lattice and discuss its
symmetries. Section \ref{sec:dimer} is devoted to the perturbative treatment in
the isolated-dimer limit. There, we first map the spin model onto an effective
spin-boson system which is well suited to our analysis. We show that the
low-energy effective Hamiltonian is related to the toric code on the honeycomb
lattice although, at lowest nontrivial order (six), one only has magnetic-like
operators. This straightforwardly implies that the low-energy excitations are
Abelian anyons with a semionic mutual statistics.  We also compute the two-spin
correlation functions for any vortex configuration up to order 6 and check
our results for two simple vortex configurations (vortex-free and vortex-full)
which allow for nonperturbative calculations. Finally, in
Sec.~\ref{sec:triangle}, we consider the isolated-triangle limit~; we show that
the effective low-energy Hamiltonian is, at lowest order (one), exactly the
Kitaev honeycomb model at the isotropic point. The next-order correction
involves three-spin interactions as well as triangular plaquette degrees of
freedom. In the vortex-free sector, this term is exactly the one studied by
Kitaev\cite{Kitaev06}, which opens a gap, and gives rise to non-Abelian
excitations.
Contrary to the isolated-dimer limit, one cannot diagonalize the effective
Hamiltonian for arbitrary vortex configurations. Thus, we focus on the
vortex-free sector and compute the fermionic gap in this limit, which is a check
of our perturbative expansion.

%
%
\section{model}
\label{sec:model}
%
%

\begin{figure}[t]
\includegraphics[width=7cm]{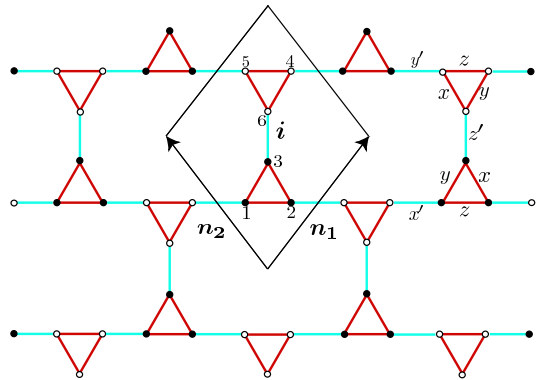}
\caption{A piece of the triangle-honeycomb lattice which has six sites per unit
  cell.}
\label{fig:Dechoney}
\end{figure}

We consider the Kitaev model on the triangle-honeycomb lattice obtained by
replacing each site of the honeycomb lattice by a triangle and described by the
following Hamiltonian~:
%
%
\begin{equation}
  \label{eq:ham}
  H=-\sum_{\alpha}\left[\sum_{\alpha-\mathrm{links}}
  J_\alpha\,\sigma_i^\alpha\sigma_j^\alpha
  +\sum_{\alpha'-\mathrm{links}}
  J'_\alpha\,\sigma_i^\alpha\sigma_j^\alpha\right],
\end{equation}
%
%
where $\alpha$ takes values $x$, $y$, or $z$, and links of type $x$, $y$, $z$
or $x^{\prime}$, $y^{\prime}$, $z^{\prime}$ are illustrated in
Fig.~\ref{fig:Dechoney}. In the above formula, $i$ and $j$ are the two sites of
the $\alpha$ or $\alpha'$ link.
Without loss of generality \cite{Kitaev06}, we also consider ferromagnetic
interactions $J_{\alpha},J'_{\alpha}>0$.
This lattice contains six sites per unit cell, and two kinds of elementary
plaquettes~: triangles and dodecagons.
As in the original Kitaev honeycomb model, $H$ commutes with all plaquette
operators defined as $W_p=\prod_{i \in p} \sigma_i^{\mathrm{out}(i)}$, where
$\mathrm{out}(i)$ denotes the outgoing direction at site $i$ with respect to the
plaquette $p$. Note that with this definition, plaquette operators have
real eigenvalues $w_p=\pm 1$, whereas, with the convention proposed in
Ref.~\onlinecite{Kitaev06}, one has $w_p=\pm \mathrm{i}$ for odd-loop operators.

As suggested in Ref.~\onlinecite{Yao07}, in the following, we further set
%
%
\begin{equation}
  J_x=J_y=J_z=J  , \quad \mathrm{and}  \quad J'_x=J'_y=J'_z=J',
\end{equation} 
%
%
so that the Hamiltonian $H$ respects the symmetries of the lattice. It is also
time-reversal invariant since it is quadratic in the spin operators. However, as
explained by Kitaev \cite{Kitaev06}, the presence of odd cycles (here due to
triangles) breaks this symmetry spontaneously. This symmetry breaking may be
understood by noting that changing the flux of all triangles, i.e.,
flipping their $W_p$'s, does not change the energy so that each eigenstate is,
at least, two-fold degenerate \cite{Yao07}.

As in the Kitaev honeycomb model \cite{Kitaev03}, the Hamiltonian $H$ can be
mapped onto a free (Majorana) fermion problem which allows for an exact
solution. Thus, in each vortex sector defined by a configuration of the $W_p$'s
one has a fermionic spectrum. Nevertheless, the low-energy states may be given
by ground states of other vortex sectors and, when the corresponding flux
configurations are not translation invariant, one is led to solve an
impurity-like problem.
In Ref.~\onlinecite{Yao07}, Yao and Kivelson numerically showed that the
ground state of $H$ always lies in the vortex-free sector ($w_p=+1$ for all $p$)
and supported this analysis by perturbative considerations \cite{Yao08}. In the
following, we shall see that this is verified in the first perturbative limit we
consider (isolated dimers), whereas we did not manage to prove it in the other
limit. In addition, in the isolated-dimer limit described in Sec.~\ref{sec:dimer}, we compute the low-energy spectrum for all vortex configurations and
give a perturbative expansion of the anyonic gap.

%
%
%
%

%
%
\section{isolated-dimer limit}
\label{sec:dimer}
%
%
%
%
\subsection{Mapping onto an effective spin-boson problem}
%
%

Our goal is to perform a perturbative analysis of the Abelian phase around the
isolated-dimer limit $J' \gg J$. To do so, we shall use the effective spin boson
mapping introduced in Ref.~\onlinecite{Schmidt08} but, for convenience, let us
first perform the following rotations~:
%
%
\begin{eqnarray}
\label{eq:rotation}
\sigma_{1, \bi}^{\alpha} &\Rightarrow&
\widetilde{\sigma}_{1, \bi}^{\,{\cal C}^2(\alpha)}, \quad 
\sigma_{2, \bi}^{\alpha} \Rightarrow
\widetilde{\sigma}_{2, \bi}^{\,{\cal C}(\alpha)}, \quad 
\sigma_{3, \bi}^{\alpha} \Rightarrow 
\widetilde{\sigma}_{3, \bi}^\alpha, \\
\sigma_{4, \bi}^{\alpha} &\Rightarrow&
\widetilde{\sigma}_{4, \bi}^{\,{\cal C}^2(\alpha)}, \quad 
\sigma_{5, \bi}^{\alpha} \Rightarrow
\widetilde{\sigma}_{5, \bi}^{\,{\cal C}(\alpha)}, \quad 
\sigma_{6, \bi}^{\alpha} \Rightarrow
\widetilde{\sigma}_{6, \bi}^\alpha, 
\end{eqnarray}
%
%
where ${\cal C}$ is the cycle operator which maps $(x,y,z)$ onto
$(y,z,x)$. Here, each site is encoded by a cell index $\bi $ and a position
index  $k=1,\dots, 6$ inside the cell as shown in  Fig.~\ref{fig:Dechoney}.
Hamiltonian (\ref{eq:ham}) then reads 
%
%
\begin{eqnarray}
\label{eq:ham2}
H&=&
-J  \sum_{\bi} 
\widetilde{\sigma}_{1, \bi}^y \widetilde{\sigma}_{2, \bi}^x+ 
\widetilde{\sigma}_{2, \bi}^y \widetilde{\sigma}_{3, \bi}^x+
\widetilde{\sigma}_{3, \bi}^y \widetilde{\sigma}_{1, \bi}^x  \nonumber \\
&&
-J  \sum_{i} 
\widetilde{\sigma}_{4, \bi}^y \widetilde{\sigma}_{5, \bi}^x+ 
\widetilde{\sigma}_{5, \bi}^y \widetilde{\sigma}_{6, \bi}^x+
\widetilde{\sigma}_{6, \bi}^y \widetilde{\sigma}_{4, \bi}^x\\
&&
-J' \sum_{i} 
\widetilde{\sigma}_{1, \bi}^z \widetilde{\sigma}_{4, \bi-\bn_1}^z+ 
\widetilde{\sigma}_{2, \bi}^z \widetilde{\sigma}_{5, \bi-\bn_2}^z+
\widetilde{\sigma}_{3, \bi}^z \widetilde{\sigma}_{6, \bi}^z. \nonumber
\end{eqnarray}
%
%
With this transformation, the interaction term on each dimer $(i,j)$ displayed
in cyan in Fig.~\ref{fig:Dechoney} is simply
$\widetilde{\sigma_i}^{z}\widetilde{\sigma_j}^{z}$. 
Denoting $|\!\uparrow\rangle$ ($|\!\downarrow\rangle$) as the eigenstate of
$\widetilde\sigma^z$ with eigenvalue of $+1$ ($-1$), each cyan dimer can be in
four different states,
%
%
\begin{equation}
\begin{array}{lll}
\{|\!\uparrow\uparrow\rangle,|\!\downarrow\downarrow\rangle\} &\mathrm{with \:\: energy}& -J' , \\
\{|\!\uparrow\downarrow\rangle,|\!\downarrow\uparrow\rangle\} &\mathrm{with \:\: energy}& +J'. 
\end{array}
\end{equation}
%
%

An alternative description of these four states consists in interpreting the
low-energy (ferro-magnetic) states as two effective spin states without a
quasiparticle and the high-energy (antiferro-magnetic) states as two effective
spin states with one quasiparticle. The energy gap between these states is
$\Delta=2J'$ and corresponds to the fermionic gap evoked in Sec.~\ref{sec:model}. In the following, we set once for all $J'=1/2$  or equivalently
$\Delta=1$.
Among the possible mappings we choose the following~: \cite{Schmidt08}
%
%
\begin{equation}
\label{eq:mappings}
|\!\uparrow\uparrow\rangle=|\!\Uparrow\! 0\rangle,
|\!\downarrow\downarrow\rangle=|\!\Downarrow\! 0\rangle,\,
|\!\uparrow\downarrow\rangle=|\!\Uparrow\! 1\rangle,
|\!\downarrow\uparrow\rangle=|\!\Downarrow\! 1\rangle,
\end{equation}
%
%
where the left (right) spin is the one of the black (white) site of the dimer,
and double arrows represent the state of the effective spin. 

Within this framework, each dimer is reduced to a single site with 4 degrees
of freedom [one effective spin $1/2$ and a hardcore boson occupation number (0
or 1)].
Considering that each site ($\bullet$ or $\circ$, see Fig.~\ref{fig:Dechoney})
of the triangle-honeycomb lattice belongs to a cyan dimer, one then has
%
%
\begin{equation}
  \label{eq:mapping}
  \begin{array}{lcl}
  \widetilde{\sigma}_{i,\bullet}^x=\tau_i^x (\crb_i+\anb_i)
  &,&
  \widetilde{\sigma}_{i,\circ}^x=\crb_i+\anb_i ,\\ 
  \widetilde{\sigma}_{i,\bullet}^y=\tau_i^y (\crb_i+\anb_i)
  &,&
  \widetilde{\sigma}_{i,\circ}^y=\mathrm{i} \, \tau^z_i (\crb_i-\anb_i) , \\
  \widetilde{\sigma}_{i,\bullet}^z=\tau_i^z
  &,&
  \widetilde{\sigma}_{i,\circ}^z=\tau_i^z (1-2 \crb_i \anb_i),
  \end{array}
\end{equation}
%
%
where ${\tau}_i^{\alpha}$'s are  Pauli matrices acting on the effective spin, and $\crb_i$ ($\anb_i$) is the creation (annihilation) operator of a hardcore boson at site $i$ which obeys
%
%
\begin{equation}
[\anb_i, \crb_j]=\delta_{ij}(1-2 \crb_i \anb_i).
\label{eq:hardcore_com}
\end{equation}
%
%

\begin{figure}[t]
\includegraphics[width=6cm]{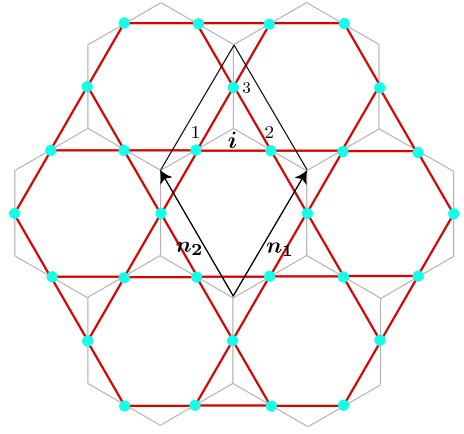}
\caption{Effective kagome lattice obtained from the triangle-honeycomb lattice by replacing each dimer linking triangles by a site. We displayed in gray the ``dual'' honeycomb lattice on which the toric code is defined (see text).}
\label{fig:kagome}
\end{figure}

Hamiltonian (\ref{eq:ham2}) then reads 
%
%
\begin{equation}
H=-\frac{N}{2}+Q+T_{0}+T_{+2}+T_{-2},
\label{eq:Hcut}
\end{equation}
%
%
where $N$ is the number of cyan dimers, $Q=\sum_{i} b_{i}^{\dagger}b_{i}$,
%
%
\begin{eqnarray}
  \label{eq:T0}
  T_0&=& -J \sum_\bi \Big( t_{1,\bi}^{2,\bi} + t_{2,\bi}^{3,\bi}
  + t_{3,\bi}^{1,\bi} \\
  &&+ t_{1,\bi+\bn_1}^{2,\bi+\bn_2} + t_{2,\bi+\bn_2}^{3,\bi}
  + t_{3,\bi}^{1,\bi+\bn_1} \Big)+ \mathrm{H.c.}, \nonumber \\
  \label{T2}
  T_{+2}=T_{-2}^{\dagger}&=& -J \sum_\bi \Big(v_{1,\bi}^{2,\bi} +
  v_{2,\bi}^{3,\bi} + v_{3,\bi}^{1,\bi} \\
  && + v_{1,\bi+\bn_1}^{2,\bi+\bn_2} + v_{2,\bi+\bn_2}^{3,\bi} +
  v_{3,\bi}^{1,\bi+\bn_1} \Big), \nonumber
\end{eqnarray}
%
%
with hopping operators,
%
%
\begin{eqnarray}
t_{1,\bi}^{2,\bi} &=& b_{2,\bi}^{\dagger}b_{1,\bi} {\tau}_{1,\bi}^{y} {\tau}_{2,\bi}^{x}, \\
t_{2,\bi}^{3,\bi} &=& b_{3,\bi}^{\dagger}b_{2,\bi} {\tau}_{2,\bi}^{y} {\tau}_{3,\bi}^{x},\\
t_{3,\bi}^{1,\bi} &=& b_{1,\bi}^{\dagger}b_{3,\bi} {\tau}_{3,\bi}^{y} {\tau}_{1,\bi}^{x}, \\
t_{1,\bi+\bn_1}^{2,\bi+\bn_2} &=& - \mathrm{i} \, b_{2,\bi+\bn_2}^{\dagger} b_{1,\bi+\bn_1} {\tau}_{1,\bi+\bn_1}^{z},\\
t_{2,\bi+\bn_2}^{3,\bi} &=& - \mathrm{i} \, b_{3,\bi}^{\dagger} b_{2,\bi+\bn_2} {\tau}_{2,\bi+\bn_2}^{z},\\
t_{3,\bi}^{1,\bi+\bn_1} &=& - \mathrm{i} \, b_{1,\bi+\bn_1}^{\dagger} b_{3,\bi} {\tau}_{3,\bi}^{z},
\end{eqnarray}
%
%
and pair-creation operators,
%
%
\begin{eqnarray}
v_{1,\bi}^{2,\bi} &=& b_{2,\bi}^{\dagger}b^{\dagger}_{1,\bi} {\tau}_{1,\bi}^{y} {\tau}_{2,\bi}^{x}, \\
v_{2,\bi}^{3,\bi} &=& b_{3,\bi}^{\dagger}b^{\dagger}_{2,\bi} {\tau}_{2,\bi}^{y} {\tau}_{3,\bi}^{x},\\
v_{3,\bi}^{1,\bi} &=& b_{1,\bi}^{\dagger}b^{\dagger}_{3,\bi} {\tau}_{3,\bi}^{y} {\tau}_{1,\bi}^{x}, \\
v_{1,\bi+\bn_1}^{2,\bi+\bn_2} &=&  \mathrm{i} \, b_{2,\bi+\bn_2}^{\dagger} b_{1,\bi+\bn_1}^{\dagger} {\tau}_{1,\bi+\bn_1}^{z},\\
v_{2,\bi+\bn_2}^{3,\bi} &=& \mathrm{i} \,            b_{3,\bi}^{\dagger}            b_{2,\bi+\bn_2}^{\dagger} {\tau}_{2,\bi+\bn_2}^{z},\\
v_{3,\bi}^{1,\bi+\bn_1} &=&  \mathrm{i} \,            b_{1,\bi+\bn_1}^{\dagger} b_{3,\bi}^{\dagger} {\tau}_{3,\bi}^{z}.
\end{eqnarray}
%
%
Now, each site is encoded by a cell index $\bi$ and its position inside the cell
which takes three values $k=1,2,3$ as shown in Fig.~\ref{fig:kagome}. Within
this formalism, the plaquette operators read
%
%
\begin{equation}
W_{{\vartriangle}}=  \prod_{i \in {\vartriangle}} \tau^z_i, \quad 
W_{{\triangledown}}=(-1)^{\sum_{i \in {\triangledown}} \crb_i \anb_i}
\prod_{i \in {\triangledown}} \tau^z_i,
\end{equation}
%
%
for triangles and 
%
%
\begin{equation}
  W_{\hexagon}= (-1)^{\crb_{2,\bi} \anb_{2,\bi}
    + \crb_{1,\bi-\bn_2} \anb_{1,\bi-\bn_2}+
    \crb_{3,\bi-\bn_1} \anb_{3,\bi-\bn_1}}
  \prod_{\bi \in {\hexagon}} \tau_i^{y},
\end{equation}
%
%
for the dodecagonal plaquette (which are hexagonal in the effective lattice) located below the cell 
$\bi$  (see notations in Fig.~\ref{fig:kagome}).

The main interest of this mapping is that the form of Hamiltonian (\ref{eq:Hcut}) is especially adapted to the perturbative treatment developed in the Sec.~\ref{subsec:spectrum}. Indeed, a key ingredient of our approach which is based on the continuous unitary transformations \cite{Wegner94} together with the particle-number conserving generator 
\cite{Stein97,Mielke98,Knetter00} is that the energy spectrum of the unperturbed Hamiltonian has to be equidistant.

%
%
%
%

%
%
\subsection{Perturbative analysis of the low-energy sector}
\label{subsec:spectrum}
%
%

In this section, we used exactly the same method as those described in
Refs.~\onlinecite{Schmidt08} and \onlinecite{Vidal08_2} for the Kitaev honeycomb model. Therefore,
we skipped all technical details and only give here the results of our
calculations.

First, let us note that, in the triangle-honeycomb lattice, one has
%
%
\begin{equation}
  N_{\mathrm{s}}=3 N_{\mathrm{t}}=6 N_{\mathrm{d}}=2 N,
\end{equation}
%
%
where $N$ is the number of cyan dimers and $\mathrm{s}$, $\mathrm{t}$, and
$\mathrm{d}$ stand, respectively for sites, triangles
(so $N_\mathrm{t}=N_\vartriangle+N_\triangledown$), and dodecagons (or effective
hexagons, so $N_{\mathrm{d}}=N_{\hexagon}$).
There are thus as many conserved $\mathbb{Z}_2$ plaquette operators
$N_{\mathrm{t}}+N_{\mathrm{d}}$ as the number $N$ of effective spin 1/2.
This implies that in the low-energy subspace with no hardcore boson, the
effective Hamiltonian can, in the isolated-dimer limit $J' \gg J$, be written
only in terms of the plaquette operators, and is thus readily solved. The
general form of this effective Hamiltonian reads
%
%
\begin{equation}
  H_\mathrm{eff}^{(0)}= E_0-\sum_{p_1,\ldots,p_n}C_{p_1,\ldots,p_n}
  W_{p_1}W_{p_2}\ldots W_{p_n},
  \label{eq:Hefflow}
\end{equation}
%
%
where ${p_1,\ldots,p_n}$ denotes a set of $n$ plaquettes.

We performed this perturbative expansion of the effective Hamiltonian up to
order $8$ which is the lowest order involving multi-plaquette interactions. At
this order, the constant term is given by
%
%
\begin{equation}
  \frac{E_0}{N}= -\frac{1}{2}-J^2-\frac{1}{4}J^4
  -\frac{1}{8}J^6+\frac{15}{64}J^8.
  \label{eq:E0}
\end{equation}
%
%
The first nontrivial contribution arises at order 6 where only
hexagonal-plaquette operators are involved.
At order 8, this latter term is renormalized and triangular-plaquette operators
come into play. More precisely, one has
%
%
\begin{equation}
  \label{eq:Heff_8}
  H_\mathrm{eff}^{(0)}= E_0-
  C_h \sum_h W_h-
  C_{h,t_1,t_2} \sum_{h,t_1,t_2} W_hW_{t_1}W_{t_2},
\end{equation}
%
%
where the first sum is performed over all hexagonal plaquettes $h$ (i.e.
${\hexagon}$) and the second one over triplet plaquettes made of one hexagon $h$
and two triangles $t_1$ and $t_2$ (of any kind $\vartriangle$ or
$\triangledown$) adjacent to this hexagon. At order 8, the coefficients are
given by 
%
%
\begin{equation}
  \label{eq:C}
  C_h= \frac{63}{8} J^6 - \frac{297}{8}J^{8}, \quad
  C_{h,t_1,t_2}=\frac{33}{16} J^8.
\end{equation}
%
%

Thus, at order 6, the spectrum does not depend on the fluxes inside the
triangles and this degeneracy is only partially lifted at order 8. Note that
the signs of $C_h$ and $C_{h,t_1,t_2}$ confirm, in this limit,
that (one of) the ground state lies in the vortex-free sector
($w_p=+1$ for all plaquettes) as conjectured in Ref.~\onlinecite{Yao07}. In
addition, we emphasize that triangular-plaquette operators appear by pairs which
are reminiscent from the time-reversal symmetry that the effective Hamiltonian
must satisfy (see Sec.~\ref{sec:model}).

It is interesting to interpret this result in terms of plaquettes and vertex
operators. Therefore, one may view the effective kagome lattice as a honeycomb
lattice where each site lies in the middle of the triangles as shown in
Fig.~\ref{fig:kagome}. Within this picture, the plaquette operators
$W_{\hexagon}$ are interpreted as flux (magnetic) operators, whereas
$W_{{\vartriangle},{\triangledown}}$ appear as vertex (electric) operators. In
this gauge theory language used in the toric-code model \cite{Kitaev03}, our
results show that, at lowest order, there is no contribution of the vertex
operators. Thus, the triangle-honeycomb lattice, in this isolated-dimer limit
does not map onto a standard toric-code-like problem.
However, the eigenstates of the effective Hamiltonian are those of the toric
code on the hexagonal lattice and, as such, display anyonic statistics. To be
more precise, one must distinguish between electric and magnetic excitations
which are localized on triangles and hexagons (in the kagome lattice),
respectively. These two kinds of excitations have mutual semionic statistics
\cite{Kitaev03,Yao07} but they individually behave as bosons. Finally, let us
remark that the gaps of magnetic excitations (order of magnitude $J^6$) and of
electric excitations (order of magnitude $J^8$) are even smaller than the gap in
the Kitaev honeycomb model (order of magnitude $J^4$). This would make an
experimental detection of anyons in the triangle-honeycomb model even more
problematic than in the honeycomb model \cite{Dusuel08_1}.

%
%
\subsection{Correlation functions in the low-energy sector}
%
%
As in the Kitaev honeycomb model, any correlator involving an odd number of spin
operators vanishes although the eigenstates break the time-reversal
symmetry. Indeed, as discussed by Yao and Kivelson \cite{Yao07}, every
eigenstate is, at least, two-fold degenerate since one can flip every triangular
plaquette without changing the energy but {\em this operation is global}.
Consequently, as in the honeycomb model \cite{Baskaran07,Chen08}, the only
nonvanishing correlators are products of $\sigma^\alpha_{i}\sigma^\alpha_{j}$ on
an $\alpha$ dimers. Here, we focus on the simplest case involving only one such
object, i.e., $C^{\alpha \alpha }_{i,j}=\langle\sigma^\alpha_{i}
\sigma^\alpha_{j}\rangle$.

In the triangle-honeycomb model, one has, {\it a priori}, nine different
functions to consider since the unit cell contains nine different
dimers. However, with the choice of the couplings we made, one only has two
different functions to distinguish : those on ``weak'' bonds ($x,y,z$ links with interaction
$J$) and those on ``strong'' bonds ($x',y',z'$ links with interaction $J'$).
As for the low-energy spectrum, one expects a plaquette-operator expansion as in
Eq.~(\ref{eq:Hefflow}).
We performed the calculation of these two correlation functions up to order 6
and obtained
%
%
\begin{equation}
  \label{eq:CFzz}
  C^\mathrm{strong}_{i,j}  = 1-2J^2 - \frac{3}{2}J^4  - \frac{5}{4}J^6
  - \frac{105}{8}J^6 \left( W_{p_1}+W_{p_2} \right),
\end{equation} 
where $p_1$ and $p_2$ are the two dodecagonal plaquettes shared by the
considered strong bond $(i,j)$.
Similarly, since we set $J_x=J_y$, we found for a weak bond $(i,j)$
%
%
\begin{equation}
  \label{eq:CFxx}
  C^\mathrm{weak}_{i,j}=J + \frac{1}{2}J^3  + \frac{3}{8}J^5
  + \frac{63}{8}J^5 W_{p},
\end{equation} 
%
%
where $p$ is the dodecagonal plaquette adjacent to the considered bond.

As can be seen from Eqs.~(\ref{eq:CFzz}) and (\ref{eq:CFxx}), the presence of a vortex is detected at orders 6 and 5, respectively. This difference stems from the fact that one analyzes the isolated-dimer limit for which, at lowest order,  $C^\mathrm{strong}_{i,j}= 1$  whereas $C^\mathrm{weak}_{i,j}=0$. 
%
%
\subsection{Checks from Majorana fermions}
%
%
To check our results we computed exactly the ground-state energy in the
vortex-free (full) sector for which  $w_p=+1$ (-1) for all $p$ using Majorana
fermions as described by Kitaev for the honeycomb model \cite{Kitaev06}.
Following, the procedure described in Ref.~\onlinecite{Vidal08_2}, we performed
a perturbative expansion of the exact solutions order by order.
Denoting  $e_0^\nu$ the ground-state energy per cyan dimer for a vortex filling
factor
$\nu=\frac{\rm Number\:\: of \:\: vortex}{\rm Number\:\: of \:\: plaquette}$ one
gets
%
%
\begin{eqnarray}
  \label{eq:gsefree}
  e_0^{\nu=0} &=& -J'-\frac{J^2}{2 J'}-\frac{J^4}{32 J'^3}
  -\frac{11J^6}{128 J'^5}+\frac{147J^8}{8192 J'^7}, \quad\\
  \label{eq:gsefull}
  e_0^{\nu=1} &=& -J'-\frac{J^2}{2 J'}-\frac{J^4}{32 J'^3}
  +\frac{5 J^6}{64 J'^5}-\frac{117J^8}{8192 J'^7}.
\end{eqnarray}
%
%
Keeping in mind that the number of hexagons is $N/3$ and that for each hexagon
$\mathrm{h}$, there are 15 triplets $\mathrm{h,t_1,t_2}$, these results are
straightforwardly recovered using Eqs.~(\ref{eq:E0})-(\ref{eq:C}).

One can also check the expression of the correlation functions in these vortex
configurations. Indeed, the Hellmann-Feynman theorem states that
%
%
\begin{eqnarray}
  \label{eq:HF_zz}
  \frac{\partial e_0^{\nu}}{\partial J'} &=&
  - \frac{1}{N}\sum_{(i,j)} C^\mathrm{strong}_{i,j}
  =- C^\mathrm{strong}_{i,j},  \\
  \label{eq:HF_xx}
  \frac{\partial e_0^{\nu}}{\partial J} &=&
  - \frac{1}{N}\sum_{(i,j)} C^\mathrm{weak}_{i,j}=
  - 2 \ C^\mathrm{weak}_{i,j},\end{eqnarray}
%
%
where the sum in Eq.~(\ref{eq:HF_zz}) [Eq.~(\ref{eq:HF_xx})] is performed over
all strong bonds (weak bonds) in the initial lattice. The last
equalities stem from the fact that for $\nu=0,1$ every plaquette has the same
contribution which would not be true for other vortex configurations.
Using Eqs.~(\ref{eq:gsefree}) and (\ref{eq:gsefull}) and the above relations, one
can easily check the validity of Eqs.~(\ref{eq:CFzz}) and (\ref{eq:CFxx}).

%
%
\section{isolated-triangle limit}
\label{sec:triangle}
%
%

\subsection{Mapping to a spin-boson plaquette problem}
%
%
\begin{figure}[t]
\includegraphics[width=7cm]{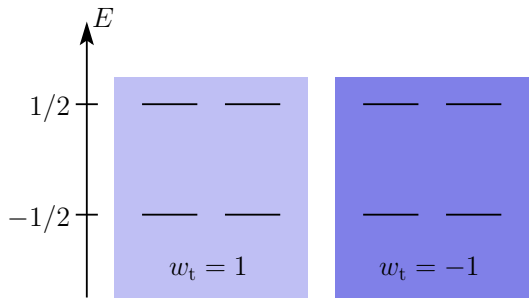}
\caption{Spectrum of an isolated triangle with $J=\frac{1}{2\sqrt{3}}$. The
  eigenstates have quantum number $w_\mathrm{t}=1$ (left) or $w_\mathrm{t}=-1$
  (right).}
\label{fig:spectrum_triangle}
\end{figure}
%
%

Let us now turn to the isolated-triangle limit $J \gg J'$. These triangles live
on the sites of an effective hexagonal lattice.
For convenience, we use rotated form (\ref{eq:ham2}) that was already used
in the isolated-dimer limit.
The spectrum of the Hamiltonian of an isolated triangle is made of two sets of
fourfold-degenerate levels. In each of these sets, two levels have eigenvalue
$w_\mathrm{t}=1$ and the other two $w_\mathrm{t}=-1$. Setting
$J=\frac{1}{2\sqrt{3}}$, the eigenenergies are $\pm 1/2$. This information is
gathered in Fig.~\ref{fig:spectrum_triangle}.

As in the isolated-dimer limit, we interpret low (high) energy states of an
isolated triangle as containing zero (one) hardcore boson. This hardcore boson
degree of freedom, which we again denote as $b$, together with the
$\mathbb{Z}_2$ quantum number $w_\mathrm{t}$ span a four-dimensional Hilbert
space. It is then natural to introduce an effective spin 1/2 to span the full
eight-dimensional Hilbert space of a triangle. We again denote this effective
spin as $\tau$ (though it is not the same as in the other perturbative limit,
and the same remark holds for $b$). The way this effective spin is introduced is
partly dictated by the operators involved in the adjacent dodecagonal-plaquette
operators. For the triangle $(1,2,3)$ of Fig.~\ref{fig:Dechoney} (and dropping the
$\bi$ index), the mapping reads
%
%
\begin{equation}
  \label{eq:tau_triangle}
  \tau^x=W_\mathrm{t}\ws_2^x\ws_3^y,\quad
  \tau^y=W_\mathrm{t}\ws_3^x\ws_1^y,\quad
  \tau^z=-W_\mathrm{t}\ws_1^x\ws_2^y.
\end{equation}
%
%
It is straightforward to check that the above $\tau^\alpha$ operators satisfy
the usual Pauli matrices algebra. The product $\ws_2^x\ws_3^y$ appearing
in $\tau^x$, for example, is the same as the one that appears in the
dodecagonal-plaquette operator having the bond $(2,3)$ in common with the
triangle.
It is interesting to note that one has to use the operator
$W_\mathrm{t}=\ws_1^z\ws_2^z\ws_3^z$ to fulfill the SU(2) algebra.
The same mapping is used for the other triangles, with $1,2$, and $3$ simply replaced
by $4,5$, and $6$ (see Fig.~\ref{fig:Dechoney} for notations).

Since the terms proportional to $J$ in Eq.~(\ref{eq:ham2}) now read
$-N_\mathrm{t}/2+Q$, with $Q=\sum_\bi \crb_\bi\anb_\bi$ where the sum runs over
the sites $\bi$ of the effective hexagonal lattice (formed by triangles), the
last task in rewriting Hamiltonian (\ref{eq:ham2}) is to find the new
form of the terms proportional to $J'$. A simple but lengthy calculation yields

%
%
\begin{equation}
  \label{eq:ham_triangle}
  H=-N_\mathrm{t}/2+Q
  -\frac{J'}{3}\sum_{\alpha=x,y,z}\sum_{\alpha-\mathrm{links}}
  \mathcal{O}_\bi^\alpha\mathcal{O}_\bj^\alpha,
\end{equation}
%
%
where $\alpha$ indicates a link of type $x$, $y$, or $z$ on the honeycomb
or equivalently brickwall lattice, with the conventions of Kitaev
\cite{Kitaev06}, as shown in Fig.~\ref{fig:brickwall}. Furthermore $\bi$ and
$\bj$ are the sites of the effective brickwall lattice on link $\alpha$, and the
operators $\mathcal{O}_\bi^\alpha$ reads ($W_{\mathrm{t},\bi}$ denotes the
triangular-plaquette operator which is now associated to site $\bi$)
%
%
\begin{equation}
  \mathcal{O}_\bi^\alpha=\tau_\bi^\alpha\Bigg\{(-1)^{\crb_\bi\anb_\bi}
  -\sqrt{2}W_{\mathrm{t},\bi}
  \left[\mathrm{e}^{\frac{2\mathrm{i}\pi}{3} p^\alpha W_{\mathrm{t},\bi}}\,
    \crb_\bi+\mathrm{H.c.}\right]\Bigg\},
  \label{eq:operatorO}  
\end{equation}
%
%
where $p^x=1, p^y=-1$, and $p^z=0$.

\subsection{Perturbation analysis of the low-energy sector}

From the above expressions, it is clear that the Hamiltonian can now be written
%
%
\begin{equation}
  H=-\frac{N_\mathrm{t}}{2}+Q+T_{0}+T_{+1}+T_{-1}+T_{+2}+T_{-2},
  \label{eq:Hcut2}
\end{equation}
%
%
where the operators $T_n$ change the number of bosons by $n$ and are
proportional to $J'$. With our notations, they read
%
\begin{equation}
  \label{eq:T012}
  T_n=-\frac{J'}{3}\sum_{\alpha=x,y,z}\sum_{\alpha-\mathrm{links}}
  \tau_\bi^\alpha\tau_\bj^\alpha\, \mathcal{T}^\alpha_{n,\bi,\bj}, 
\end{equation}  
%
%
with
%
%
\begin{eqnarray}
  \mathcal{T}^\alpha_{0,\bi,\bj}&=&(-1)^{\crb_\bi\anb_\bi+\crb_\bj\anb_\bj}\\
  &&+2\,W_{\mathrm{t},\bi}W_{\mathrm{t},\bj}\left[
    \mathrm{e}^{\frac{2\mathrm{i}\pi}{3} p^\alpha
      (W_{\mathrm{t},\bi}-W_{\mathrm{t},\bj})}\, \crb_\bi\anb_\bj
    +\mathrm{H.c.}\right],\nonumber\\
  \mathcal{T}^\alpha_{1,\bi,\bj}&=&-\sqrt{2}\Big[W_{\mathrm{t},\bi}\,
  \mathrm{e}^{\frac{2\mathrm{i}\pi}{3} p^\alpha W_{\mathrm{t},\bi}}\,
  \crb_\bi\,(-1)^{\crb_\bj\anb_\bj}+(\bi\leftrightarrow\bj)\Big]\nonumber\\
  &&\\
  \mathcal{T}^\alpha_{2,\bi,\bj}&=&2\,W_{\mathrm{t},\bi}W_{\mathrm{t},\bj}\,
  \mathrm{e}^{\frac{2\mathrm{i}\pi}{3} p^\alpha
    (W_{\mathrm{t},\bi}+W_{\mathrm{t},\bj})}\, \crb_\bi\crb_\bj,\\
  &&T_{-1}=T_1^\dagger, \quad \mathrm{and}\quad T_{-2}=T_2^\dagger.
\end{eqnarray}
%
%
%
%
\begin{figure}[t]
  \includegraphics[width=0.9\linewidth]{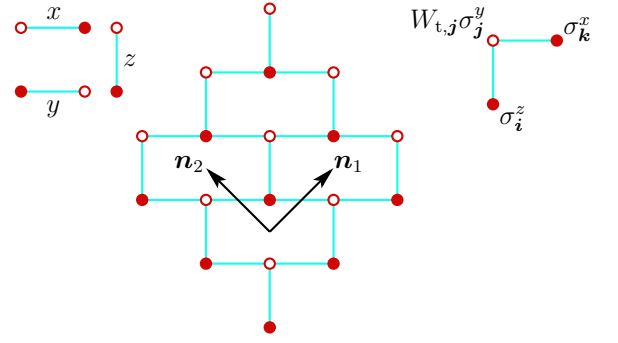}
  \caption{A piece of the effective brickwall lattice spanned by triangles,
    together with the notations for $x$, $y$, and $z$ links (left).
    Filled (empty) dots represent the triangles made of filled (empty) dots in
    Fig.~\ref{fig:Dechoney}. On the right, we have represented one of the
    second-order three-spin terms that appear in Eq.~(\ref{eq:Hefflow2}).}
  \label{fig:brickwall}
\end{figure}
%
%

As in the isolated-dimer limit (where $T_{\pm 1}$ terms are absent), with a
suitable unitary transformation, such a Hamiltonian can be recasted in a
unitary equivalent effective form $H_\mathrm{eff}$ which conserves the number of
bosons, i.e., that commutes with $Q$. We refer the reader to
Ref.~\onlinecite{Knetter00} (especially Appendix B), from which it follows that
at order 2,
%
%
\begin{equation}
  H_\mathrm{eff}=-\frac{N_\mathrm{t}}{2}+Q+T_{0}+\left[T_{+1},T_{-1}\right]
  +\frac{1}{2}\left[T_{+2},T_{-2}\right].
  \label{eq:Heff12}
\end{equation}
%
%

A tedious calculation then shows that in the low-energy subspace with no boson,
the effective Hamiltonian (still at order 2) has the form
%
%
\begin{eqnarray}
  H_\mathrm{eff}^{(0)}&=&-\frac{N_\mathrm{t}}{2}\left(1+2{J'}^2\right)
  -\frac{J'}{3}\sum_\alpha\sum_{\alpha-\mathrm{links}}
  \tau_\bi^\alpha\tau_\bj^\alpha\nonumber\\
  &&+\frac{2\sqrt{3}{J'}^2}{9}{\sum_{\bi,\bj,\bk}}'
  W_{\mathrm{t},\bj}\,\tau_\bi^\alpha\tau_\bj^\beta\tau_\bk^\gamma.
  \label{eq:Hefflow2}
\end{eqnarray}
%
%
The low-energy effective Hamiltonian at order 1 is nothing but that of the
Kitaev honeycomb model at the isotropic point, though one should remember that
site $\bi$ now also has the $\mathbb{Z}_2$ supplementary degree of freedom
$W_{\mathrm{t},\bi}$.
In Eq.~(\ref{eq:Hefflow2}), the last (primed) sum is to be taken over all possible
combinations of three sites $\bi$, $\bj$, and $\bk$ such that $\bi$ and $\bk$ are
nearest neighbors of $\bj$, and the spin ``directions'' $\alpha$, $\beta$, and
$\gamma$ are such that $(\bi,\bj)$ is an $\alpha$ link, $(\bj,\bk)$ is a
$\gamma$ link, and $\beta$ is the outgoing direction at site $\bj$ of the path
$(\bi,\bj,\bk)$ (note that $\alpha$, $\beta$ and $\gamma$ are all distinct). For
clarity, one such term is illustrated in Fig.~\ref{fig:brickwall} (right). These
terms, apart from the plaquette operator $W_{\mathrm{t},\bj}$, are exactly the
ones that arise when switching a magnetic field on, in the gapless phase of the
Kitaev honeycomb model. They open a gap and give proper non-Abelian anyonic
statistics to the vortices, as detailed in Ref.~\onlinecite{Kitaev06}.
Actually, one does not know how to diagonalize analytically  $H_\mathrm{eff}^{(0)}$ for arbitrary vortex configurations  even at order 1 \cite{Kitaev06} and, in particular, how to obtain the ground state of each sector. Therefore, contrary to the isolated-dimer limit, one cannot compute the correlation functions in this limit. 

Let us remark that the plaquette operators on an elementary brick (or hexagon)
$h$, namely $W_h=\prod_{\bi \in h} \tau_\bi^{\mathrm{out}(\bi)}$ are the product
of the plaquette operators of the corresponding dodecagon on the original
lattice {\em and} of its adjacent triangles, as follows from
Eq.~(\ref{eq:tau_triangle}).
From Eq.~(\ref{eq:Hefflow2}) and the previous remarks, it follows that the
vortex-free sector contains non-Abelian anyons, which is consistent with the
findings of Ref.~\onlinecite{Yao07}. One can also use Kitaev's result (see
Sec.~8 of Ref.~\onlinecite{Kitaev06}) to obtain the gap in this sector,
%
%
\begin{equation}
  \Delta=6\sqrt{3}\times\frac{2\sqrt{3}{J'}^2}{9}=4{J'}^2,
  \label{eq:gap}
\end{equation}
%
%
in units where $J=\frac{1}{2\sqrt{3}}$, and thus also
$\Delta=\frac{2\sqrt{3}{J'}^2}{3J}$ if $J$ is chosen freely. This value of the
gap is consistent with the numerical results obtained by Yao and
Kivelson\cite{Yao07}, and we also checked it using an expansion of the exact
result from Majorana fermionization. We also checked that the ground-state
energies in the vortex-free sector obtained, thanks to a Majorana
fermionization of Hamiltonian (\ref{eq:Hefflow2}) or directly of initial
Hamiltonian (\ref{eq:ham}), match at order 2.

Finally, let us mention that at order 1, the ground states of
Eq.~(\ref{eq:Hefflow2}) are such that $W_h=1$ for all $h$. There are many such
states. The vortex-free state is such a state, but any configuration where the
six triangles surrounding one dodecagon are flipped to $W_\mathrm{t}=-1$ is also
such a state, since every dodecagon is then surrounded by an even number of
flipped triangles. From form (\ref{eq:Hefflow2}) at order 2, we do not know
how to prove that the ground state is the vortex-free state (we do not even know
if this is true or if one has to go to higher orders in perturbation to prove
it). We leave this as an open question.

%
%
\section{Conclusion}
\label{sec:conclusion}
%
%

In this work, we have studied perturbatively the Kitaev model on the
triangle-honeycomb model. This has allowed us to show that in the isolated-dimer
limit, the model has low-energy Abelian anyonic excitations, whereas in the
isolated-triangle limit, the anyons become non-Abelian. This picture is
consistent with the values of the Chern number in each of these phases
\cite{Yao07}. In the isolated-dimer limit, we have furthermore computed the
low-energy spectrum, as well as the spin-spin correlation functions, which both
display a plaquette expansion. We emphasize that such a computation is not an
easy task within the Majorana or Jordan-Wigner formalism, which are only
well suited to study analytically configurations of vortices which display
translational invariance.

\acknowledgments

We thank H. Yao and S.~A. Kivelson for fruitful and stimulating discussions.
K.~P.~S. acknowledges ESF and EuroHorcs for funding through EURYI.


\end{document}